\documentstyle[aps]{revtex}
\begin{document}
\title{Order parameter of A-like phase of $^3$He in aerogel.}
\author {I.A.Fomin}
\address{P. L. Kapitza Institute for Physical Problems,
\\Kosygina 2, 119334 Moscow, Russia}


\maketitle
\begin{abstract}
 Phenomenological criterion of a choice of the order parameter
 for superfluid phases of $^3$He  in aerogel in a near vicinity of
 the transition temperature is formulated.
 Except for the  BW-phase this criterion is met by the order parameter
 of axiplanar phase for a special choice of its parameters.
 It is proposed as a limiting at $T\to T_c$  form of the order parameter
 for the observed A-like phase.

 \end{abstract}
\bigskip

\section {}
 Two superfluid phases of $^3$He in aerogel are usually referred as A-like
 and B-like. This reflects both their relation to the phases in the pure
 (free of impurities)  $^3$He and the existing uncertainty in their
 identification. The pulse-NMR experiments \cite{dmit} indicate strongly that
 in the B-like phase the order parameter
 after averaging over small scale fluctuations  has the same BW-form
 as that in the B-phase of a pure $^3$He.
 The measured magnetic susceptibility of A-like phase is the same as in the
normal phase \cite{osher}. This is an evidence that the order parameter in the A-like phase has an equal spin pairing (ESP) form i.e. there are no Cooper pairs with zero spin projection on a direction of magnetic field ($z$-axis)
Formally it means that the order parameter -- matrix $A_{\mu j}$ can be
represented as
$$
A_{\mu j}=\hat x_{\mu} a_j+\hat y_{\mu} b_j , \eqno(1)
$$
where $\hat x_{\mu}$ and $\hat y_{\mu}$ are the orts of the corresponding axes
 in spin space, $a_j$ and $b_j$ are complex vectors in momentum space. A-phase of
pure $^3$He (axial) is a particular case of (1). Its order parameter has a form:
$$
A_{\mu j}=\Delta\frac{1}{\sqrt{2}}\hat d_{\mu}(\hat m_j+i\hat n_j), \eqno(2)
$$
i.e. it contains only one spin vector $\hat d_{\mu}$.
The "orbital" vectors  $\hat m_j$ and $\hat n_j$ are real and orthonormal.
Volovik has shown \cite{volovik}, that  aerogel destroys the long range ordering described by the order parameter of a form, given by  eq. (2). Direction of the vector ${\bf\hat l}={\bf\hat m}\times{\bf\hat n}$  experiences a random walk and an average
of  $A_{\mu j}$ turns to zero.
There remains possibility of transition in a superfluid glass state \cite{volov1} with the order parameter formed by average of four creation and (or)
annihilation operators.
It will be shown in what follows that there exist a possibility to preserve
long range ordering for a p-wave Cooper paired phase in aerogel with the order
parameter of ESP type. Definite form of the order parameter is suggested  for the A-like
phase of  $^3$He in a near vicinity of $T_c$.

\section {}
 In a vicinity of $T_c$ interaction of aerogel with the superfluid  $^3$He 
can be described phenomenologically  \cite{fom} by adding to the Ginzburg and Landau functional an energy density of a form:  
$$
f_{\eta}=g_{\eta}\eta_{j l}({\bf r})A_{\mu j}A_{\mu l}^* ,        \eqno(3)
$$
where  $\eta_{j l}({\bf r}) $ -- is a random real symmetric tensor, $g_{\eta}$ 
-- an interaction constant. This term takes into account fluctuations in position and configurations of strands. The functional takes the form:
$$
F_{GL}=\int d^3r\{\alpha(T-T_c)A_{\mu j}A_{\mu j}^*+f_{\eta}+
      f_{\nabla}+f_4\},                                           \eqno(4)
$$
where $f_{\nabla}$ and $f_4$ are correspondingly the gradient energy and the 
fourth order terms. The isotropic part of the added tensor  
 $\eta_0({\bf r})\delta_{jl}$ can be absorbed in the  
 $T_c$, then  $\eta_{j j}=0$. Aerogel is assumed to be isotropic on the average i.e.  $<\eta_{j l}({\bf r})>=0 $, but locally it introduces deviations from spherical symmetry. 
Tensor  $\eta_{j l}({\bf r}) $ describes the splitting of  $T_c$ 
due to these deviations. In a pure  $^3$He  $T_c$ is the same for all spherical harmonics with  $l=1$. In aerogel in volumes with a size  $>\xi_0$ one can speak of a ``local transition temperature'' which is different for different projection of angular momentum. The aerogel $^3$He-interaction  $f_{\eta}$ -- is of 
the second order on $A_{\mu j}$. If $<A_{\mu j}>\ne$0,   in a close vicinity of   $T_c$ this term  gives a principal order contribution to the functional (4).
 In that region the proper choice of combinations of spherical harmonics 
 forming  $A_{\mu j}$ is determined by the perturbation  $f_{\eta}$, as in a degenerate case of perturbation theory. 
Of particular interest are combinations of spin and angular momentum projections, which are not splitted by the field   $\eta_{j l}({\bf r}) $, 
i.e. which meet the condition 
  $$
   \eta_{j l}A_{\mu j}A_{\mu l}^*=0.                              \eqno(5)
  $$
In that case $f_{\eta}$ is not a dominating contribution and the other terms, driving the phase transition  become essential.  
If  $A_{\mu j}$ is an extremum of the functional (4) and simultaneously satisfies eq. (5), then  
 $\frac{\partial F_{GL}}{\partial g_{\eta}}=0$ and a change of the interaction with aerogel does not change the energy of $^3$He. This is the case for the BW-phase: 
 $A_{\mu j} = \Delta e^{i\varphi}R_{\mu j}$, where $R_{\mu j}$- orthogonal 
matrix. It can be checked by  direct substitution in eq. (2): 
$$
 \eta_{j l}R_{\mu j}R_{\mu l}=\eta_{j l}\delta_{j l}=\eta_{j j}=0.\eqno(6)
$$
For the axial, phase going through the same argument one arrives at a finite contribution to the energy functional:
$f_{\eta}\sim -\eta_{jn}l_jl_n$ which results in the loss of orientational long range order. Let us find out whether condition (5) can be satisfied by an ESP-type order parameter. For that we write explicitly  real and imaginary parts of the vectors  {\bf a} and {\bf b} in the definition (1): 
{\bf a} ={\bf m}+i{\bf n}, {\bf b}={\bf l}+i{\bf p}, where 
{\bf m},{\bf n},{\bf l},{\bf p} -- are real vectors, and substitute eq.(1) into eq. (5).  The imaginary part of the resulting expression turns to zero automatically since $\eta_{j l}$ is symmetric.  The real part is zero if the following equation is satisfied: 
$$
 m_jm_l+n_jn_l+l_jl_l+p_jp_l=\delta_{jl}\cdot const.                  \eqno(7)
$$
The constant in the r.h.s. can be set to unity by normalization. Eq. (7) has following solutions: one of the four vectors, for example  
${\bf p}=0$, Three remaining vectors {\bf m,n,l} form an orthonormal set. As a result the order parameter has the form:  
$$
A_{\mu j}=\Delta\frac{1}{\sqrt{3}}[\hat d_{\mu}(\hat m_j+i\hat n_j)+
        \hat e_{\mu}\hat l_j].                                        \eqno(8)
$$
One can check directly that  matrix (8) satisfies eq. (5). It need not be an extremum of the functional of free energy for pure  $^3$He 
 \cite{march}, nevertheless the order parameter which is  found, is a particular case of axiplanar phase  \cite{mermin,hall} , its order parameter is proportional to 
$$
(\hat{\bf d}+i\hat{\bf e})[\hat{\bf m}v_x+i(\hat{\bf n}v_y+\hat{\bf l}v_z)]+
(\hat{\bf d}-i\hat{\bf e})[\hat{\bf m}v_x+i(\hat{\bf n}v_y-\hat{\bf l}v_z)].
                                                                   \eqno(9)
$$
This expression contains three real parameters 
$v_x, v_y, v_z$, bound by a condition: $v_x^2+v_y^2+v_z^2=1$. The isotropic ESP-phase (8) is obtained at  $v_x^2=v_y^2=v_z^2=1/3$, and the axial phase (2) -- 
at $v_x^2=v_y^2=1/2, v_z=0$. 
Both limiting cases belong to a one-parametric family 
$v_x=v_y\equiv u, v_z\equiv w, 2u^2+w^2=1$ with the order parameter proportional to 
$$
u\hat{\bf d}(\hat{\bf m}+i\hat{\bf n})-w\hat{\bf e}\hat{\bf l}     \eqno(10)
$$
When $w\ne 0$ this order parameter corresponds to a non unitary phase with a 
symmetry which is different from that of the axial phase. In particular, the  symmetry with respect to combination of a gauge transformation with the rotation of  
${\bf\hat m}$ and ${\bf\hat n}$ around ${\bf\hat l}$  is absent.

\section {}
According to the above argument in aerogel a phase transition characterized by 3 by 3 matrix order parameter $A_{\mu j}$ at   $T=T_c$ can take place either in the BW-phase or in the symmetric ESP-phase (8). In a magnetic field the phase (8) is favored by its greater magnetic susceptibility. When temperature is lowered, the coefficients    $u$ and  $w$ can deviate from the value $u=w=1/\sqrt 3$. To estimate a distance from  $T_c$, for substantial deviations to occure one has to extrapolate fluctuational corrections to 
the average value of   $A_{\mu j}$ from a region far from $T_c$, where the corrections are small into a region closer to $T_c$ where the corrections become of the order of 1.  
 In conventional superconductors \cite{Larkin} this happens at 
$(T_c-T)/T_c\sim\left(\frac{\lambda_{corr}^3}{l_{tr}^2\xi_0}\right)^2$.
Here  $\lambda_{corr}$ is a correlation length for the field 
$\eta_{j l}({\bf r})$, $l_{tr}$ --a transport mean free path for Fermi excitations, $\xi_0$ -- the superfluid coherence length in the liquid  $^3$He. Assuming 
$\lambda_{corr}\sim$500\AA, $l_{tr}\sim$2000\AA, $\xi_0\sim$200\AA, one arrives at $(T_c-T)/T_c\sim$1/30.  This estimation is not very accurate because poorly known  $\lambda_{corr}$ enters it in the sixth power. More reliable estimation one can extract from the data on the smearing of the specific heat jump 
\cite{parpia}. According to this data  $(T_c-T)/T_c\sim$1/25.
At a smaller deviation from $T_c$ the order parameter must be close to the symmetric ESP-phase. This region can be even larger if the phase in question is close to a stable minimum  of the functional (4) for  $f_{\eta}$=0 and for the existing 
 $f_4$. The symmetric ESP-phase does not experience the  disorienting effect of aerogel. When 
 $u\ne w$ the disorienting term is finite
$(w^2-u^2)\eta_{j l}l_jl_l$, but if 
 $|w^2-u^2|\ll$1 the corresponding correlation length  \cite{volovik} is much greater then the dipole length. In that case directions of {\bf l} and {\bf m} 
are fixed by  {\bf d} and {\bf e}.
To find a limit of stability of the symmetric ESP-phase one needs a detailed quantitative analysis, which would involve poorly known parameters. 

Possible experiments which would make distinction between the axial and axiplanar phases  were discussed in a literature and performed in a pure $^3$He \cite{hall}.
All of them can, in principle, be applied to helium in aerogel. More direct are measurements of the orbital properties, like anisotropy of the superfluid density. In the phase (8) the superfluid density must be isotropic.
Transverse NMR frequency shift for an equilibrium configuration in the axiplanar phase must be positive in a contrast to the observed negative shift in the A-like phase \cite{osher}. This contradiction does not exclude an identification of that phase which was proposed here. One knows that in non-equilibrium configurations one can have negative shift. It happens in the A-phase  when  {\bf d} deviates  from {\bf l} because of a presence of longitudinal oscillations \cite{fomin} or because of an  orienting effect of the walls \cite{char}. 

The author acknowledges stimulating discussions with V.V.Dmitriev, 
V.I.Marchenko  and T.E.Panov. This work was supported in part  by  
 CRDF, through research grant RP1-2089 and by RFBR under  project 01-02-16714.

  \end{document}